# Treatment of an accelerating observer via the special theory of relativity

David A. de Wolf *
*Virginia Tech, Blacksburg, VA, 24061*
*dadewolf340723@gmail.com*

## Abstract

The clock time $t'$ of an accelerating observer, simultaneous with the measured clock time $t$ of an inertial observer, is easily established in special relativity (as proper time) by the well-known time-dilation formula for $t'(t)$. In this work, a special-relativity formula $t(t')$ for an initial observer's clock time $t$, simultaneous with a traveler's measured clock time $t'$, is derived and applied successfully to various relevant situations. The equation for $t(t')$ requires knowledge only of the velocity profile $v(t')$ of the traveler.

## I. INTRODUCTION

It is well-known that the proper time $t'$ of an accelerating observer $\mathfrak{B}$ is expressed as a function of an inertial-frame observer's ($\mathfrak{A}$'s) time $t$ by[1]

$$t' = \int_0^{t'} dt[1-\beta^2(t)]^{1/2} \tag{1}$$

given that $\beta(t) = v(t)/c$ in terms of $\mathfrak{B}$'s velocity $v(t)$ as observed by $\mathfrak{A}$. Proper time $t'$ represents the inferred clock time of $\mathfrak{B}$ simultaneous with the measured inertial-frame clock time $t$ of $\mathfrak{A}$. The 'reciprocal' problem, expression of inferred simultaneous time $t$ of the inertial-frame observer $\mathfrak{A}$ in terms of the measured time $t'$ by $\mathfrak{B}$ is the subject of this work. Acceleration of the spacetime frame of $\mathfrak{B}$ with respect to the inertial frame makes it not possible to employ (1) with reversal of $t$ and $t'$ because (1) requires integration in an inertial frame. Therefore, a special theory-of-relativity (STR) formula, Eq. (5), for $t(t')$ that has general applicability is derived in terms of $\beta(t')$ in section II. Consequent sections give applications of Eq. (5) for several physical examples.

In the standard version of the twin paradox briefly discussed in section III, $\mathfrak{B}$ travels at constant velocity to a distant star (reaching it at $t' = T'$) and back. Traveler $\mathfrak{B}$'s velocity is constant during the entire trip, except at the star when inertial-frame observer $\mathfrak{A}$ appears to age suddenly and discontinuously at $t' = T'$ according to $\mathfrak{B}$. At that moment $\mathfrak{B}$ has reversed travel direction and has moved from one into another inertial frame. This effect is documented in very many papers; those listed below[2-6] are just a sample.

In section IV a more realistic velocity profile for $\mathfrak{B}$'s motion is introduced to demonstrate with Eq. (5) that the actual aging of $\mathfrak{A}$ changes smoothly and continuously instead of making a sudden discontinuous jump at $t' = T'$ as is the case in section III.

Finally, the formula $t(t')$of Eq. (5) is applied to the case in which $\mathfrak{B}$ experiences gravitational acceleration $g(t')$. Specifically, a result first due to Perrin[7] is obtained from Eq. (5). Where deemed helpful, appendices provide mathematical details.

## II. GENERAL STR FORMULA FOR $t(t')$

The key to finding a formula supplanting (1) is the concept of simultaneity in STR. It is helpful for that purpose to consider Fig. 1:

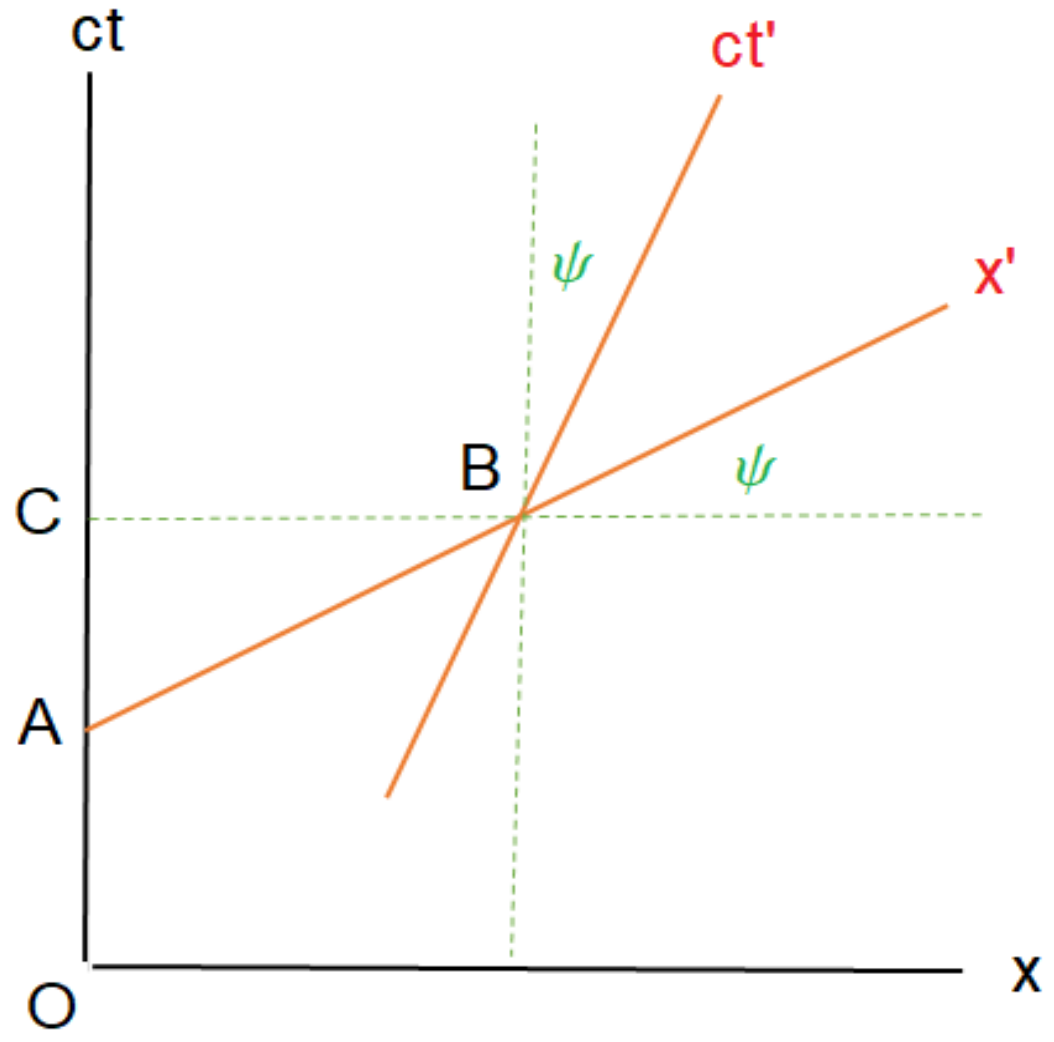

Fig. 1: Diagram to illustrate simultaneity

Consider point $B$ on $\mathfrak{B}$'s spacetime trajectory. Given that $\mathfrak{B}$'s normalized velocity is $\beta(t') = v(t')/c$ at $B$, the Lorentz-transformed coordinate frame with axes $x'$, $ct'$ is given according to Minkowski (see footnote 20 of ref. 8) by the indicated axes where $\beta(t') = \tanh\psi(t')$. Crucial in this diagram is that events simultaneous for observer $\mathfrak{A}$ lie on the line $BC$, but events simultaneous for observer $\mathfrak{B}$ lie on the line $AB$. This difference in what each observer sees as a simultaneous event was first formulated by Einstein as part of his special theory of relativity[9].

A similar formula for $t = t_A$ in terms of $t' = t_B$ (see also Figs. 2), is more complicated because 𝔅's trajectory does not lie in an inertial frame. For example, a discontinuous change of $\beta(t)$ in Eq. (1) yields a continuous change in $t'$. In $dt = dt'[1-\beta^2(t')]^{1/2}$ in contrast, a discontinuous change in $\beta(t')$ produces a discontinuous shift in point $A$ and thus in $dt$. Partly for that reason, Eq. (1) with $t$ and $t'$ interchanged is not valid.

To replace Eq. (1) for the reciprocal situation consider Fig. 2:

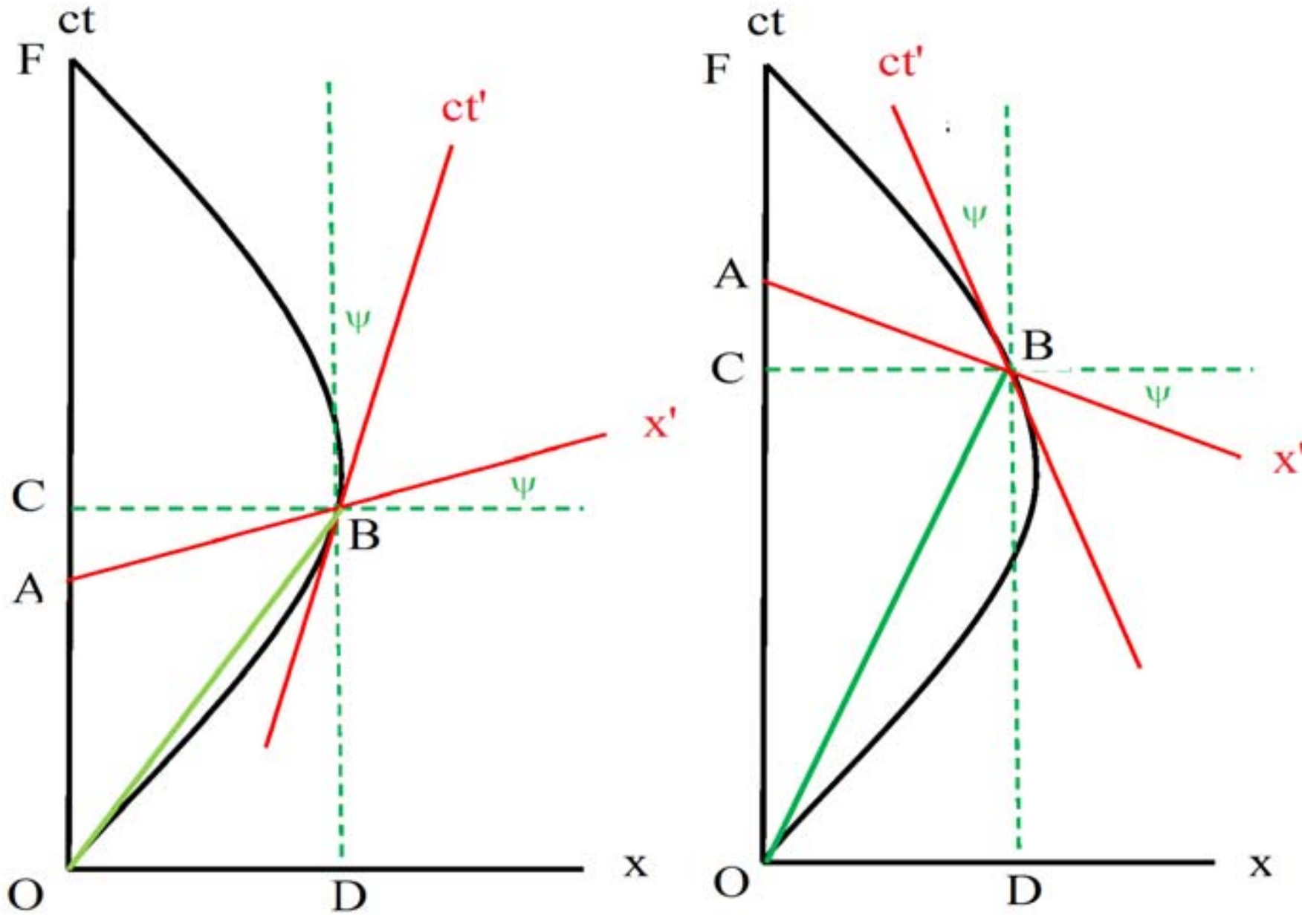

Figs. 2. Trajectory of 𝔅 in 𝔄's coordinate frame $S$ with Minkowski coordinates at point $B$.

Figures 2a and 2b, which are identical except for the location of point $B$, illustrate the trajectory of 𝔅 for a (symmetric) round trip to a star in more detail, as viewed in the inertial $(ct, x)$ frame of 𝔄 who is earthbound. All effects of Earth's gravity are ignored in these figures. Figure 2a is for $0 < t < T$ whereas Fig. 2b is for $T < t < 2T$ (return trip). The discontinuous jump of point $A$ is obvious if $OB$ and $BF$ are straight lines.

Noting again that points $A$ and $B$ are simultaneous for observer 𝔅 it follows from the hyperbolic geometry in Fig. 1a for $0 < t' < T'/2$ that

$$cdt_A \equiv d(OA) = d(OC) - d(AC) = cdt_C - d[(BC)\tanh\psi].$$

$$= cdt_B(1-\beta^2)^{-1/2} - d[\beta(OD)] = cdt_B(1-\beta^2)^{-1/2} - d(\beta x_D) \quad (1)$$

Since all terms are expressed as infinitesimals it follows that

$$ct_A = c\int_0^{t_B} dt'[1-\beta^2(t')]^{-1/2} - \beta(t_B)x_D \quad (2)$$

There is an apparent inconsistency here in using $\beta$ as given by $\mathfrak{A}$, but expressed as a function of $t'$, which will be dealt with below. Distance $x_D$ has been covered by $\mathfrak{A}$ in time $t_C$ and is therefore equal to

$$x_D = \int_0^{t_C} dt v(t) = c\int_0^{t_C} dt\beta(t) = c\int_0^{t_B} dt' \frac{\beta(t')}{\sqrt{1-\beta^2(t')}} \quad (3)$$

It is necessary in (3) to have replaced $dt$ by $dt'\cosh\psi$ in the integrand, and then $t_C$ in the upper bound by $t_B \equiv t'$ (which is simultaneous with $t_C$ in $\mathfrak{A}$'s frame, hence is identical with it). Velocity $\beta(t') = \beta(t)$ where it is understood that the value at point $C$ is needed. The result[10] is

$$t = \int_0^{t'} dt_1[1-\beta^2(t_1)]^{-1/2} - \beta(t')\int_0^{t'} dt_1 \frac{\beta(t_1)}{\sqrt{1-\beta^2(t_1)}} \quad (4)$$

When $t' > T'/2$, $d(OC) - d(AC)$ becomes $d(OC) + d(AC)$ in (1) because $\beta(t') < 0$. Equation (4) suffices also in this case due to the sign change in $\beta(t')$, which indicates that the situation of Fig. 2b applies. Lest it be thought that $\mathfrak{B}$ needs to know $v(t)$ as registered by $\mathfrak{A}$, let it be clarified here that $\beta(t_1) = \tanh(gt'_1/c)$ as can be verified from Eqs. (A2) and (A3), and time $t'_1$ as well as acceleration $g$ are therefore directly measurable by $\mathfrak{B}$. Finally, this relationship between $t$ and $t'$ holds for an arbitrary velocity distribution of the accelerating observer.

## III. STANDARD TWIN-PARADOX TIME DILATION RESULT

It is trivially verifiable that the standard twin-paradox (TP) results[2] when $\beta(t) = \beta$ (a constant) for $t < T$ and $\beta(t) = -\beta$ for $T < t < 2T$. Let $\gamma \equiv 1/\sqrt{1-\beta^2}$. Then Eq. (5) yields

$$t = t'\sqrt{1-\beta^2} \quad \text{for } t' < T'$$
$$t = t'(1+\beta^2)/\sqrt{1-\beta^2} \quad \text{for } T' < t' < 2T'$$

$$T = T'/\sqrt{1-\beta^2} \qquad (5)$$

$\mathfrak{A}$'s clock time $t$ ($\equiv t_A$) registered when $\mathfrak{B}$'s clock time is $t'$ ($\equiv t_B$) makes a discontinuous jump when $t' = T'$. The jump is from $t = T(1-\beta^2)$ to $t = T(1+\beta^2)$, and $t = T$ is the midpoint of the discontinuity in $t$. It is shown as calculated from (6) in Fig. 3. Also obvious from the graph, $\mathfrak{A}$'s clock advances more slowly than $\mathfrak{B}$'s everywhere except at $t' = T'$, but the discontinuous increase at $t' = T'$ more than compensates for that.

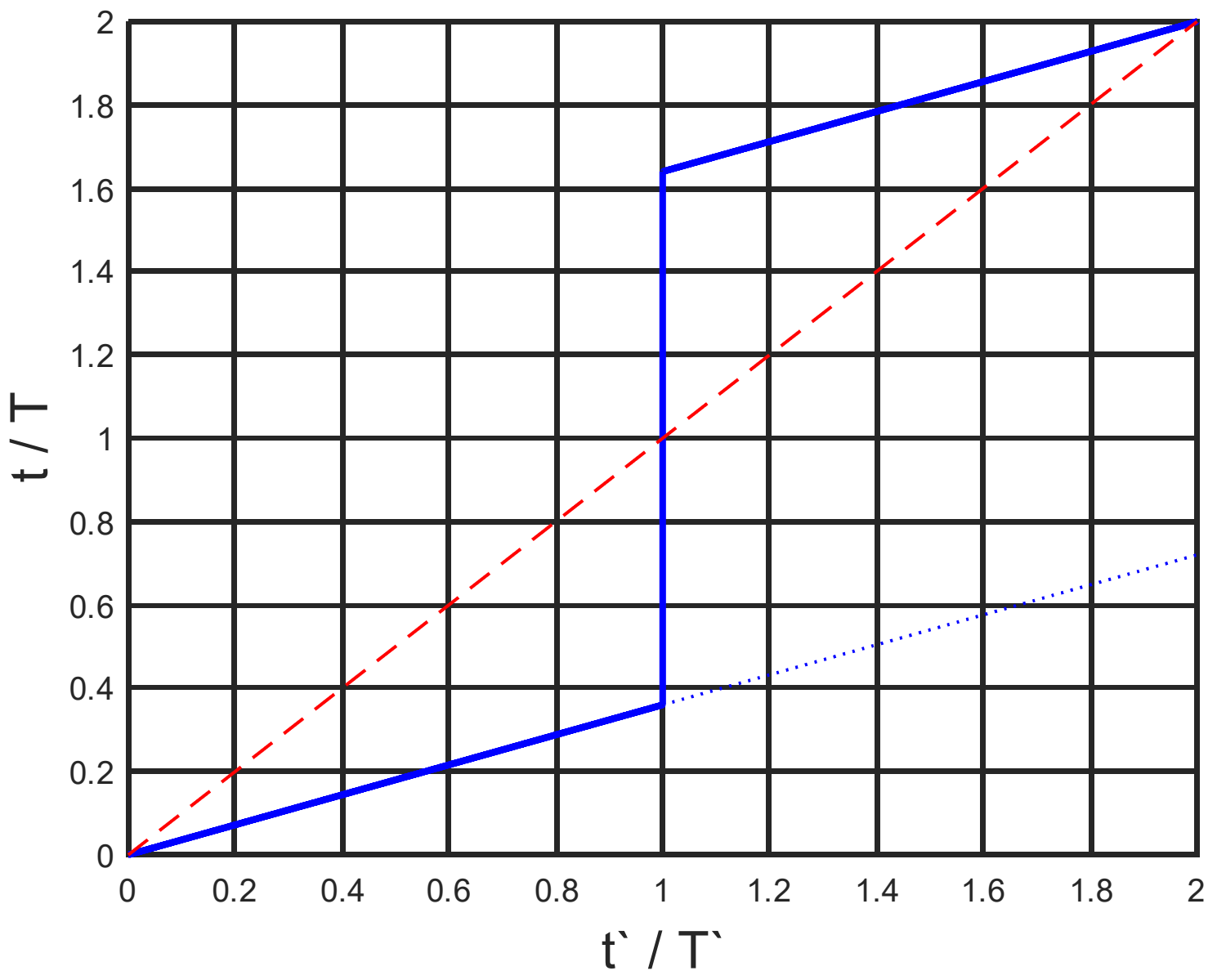

Fig. 3. $\mathfrak{A}$'s aging time $t = t_A$ vs. $t' = t_B$ for the standard TP ($\beta = 0.8$); dashed curve shows $t/T = t'/T'$ as a reference.

In any more realistic version of the TP, the transition from $v = +\beta$ to $-\beta$ is not discontinuous; it should be smooth and continuous. This will be shown explicitly for a sinusoidal velocity distribution (see section IV). While the time $t(t')$ is $\mathfrak{A}$'s actual clock time at time $t' = t_B$ for $\mathfrak{B}$, $\mathfrak{A}$ does not experience such a discontinuity, furthermore $\mathfrak{A}$ predicts simultaneous time $t'$on $\mathfrak{B}$'s clock to be given by (1), which would show, for $t' = T'$ to $t' = 2T'$ in Fig. 3, as a dotted-line continuation. Plots for other values of $\beta$ are similar.

## IV. A SINUSOIDAL VELOCITY PROFILE WITH ZERO END VELOCITIES AND ACCELERATIONS

The velocity profile (7) for a 'twin-paradox' trip is more realistic in that both velocity and acceleration are zero at $t = 0,\ T,$ and $2T$ without discontinuities:

$$\beta_1(t) = \tfrac{1}{2}\beta_0[1 - \cos(2\pi t/T)] \qquad \text{for } 0 < t < T$$
$$\beta_2(t) = -\tfrac{1}{2}\beta_0[1 - \cos(2\pi t/T)] = -\beta_1(t) \qquad \text{for } T < t < 2T \tag{6}$$

and it is shown in Fig. 4 for $\beta_0 = 0.8$ (similar profiles for other values of $\beta$):

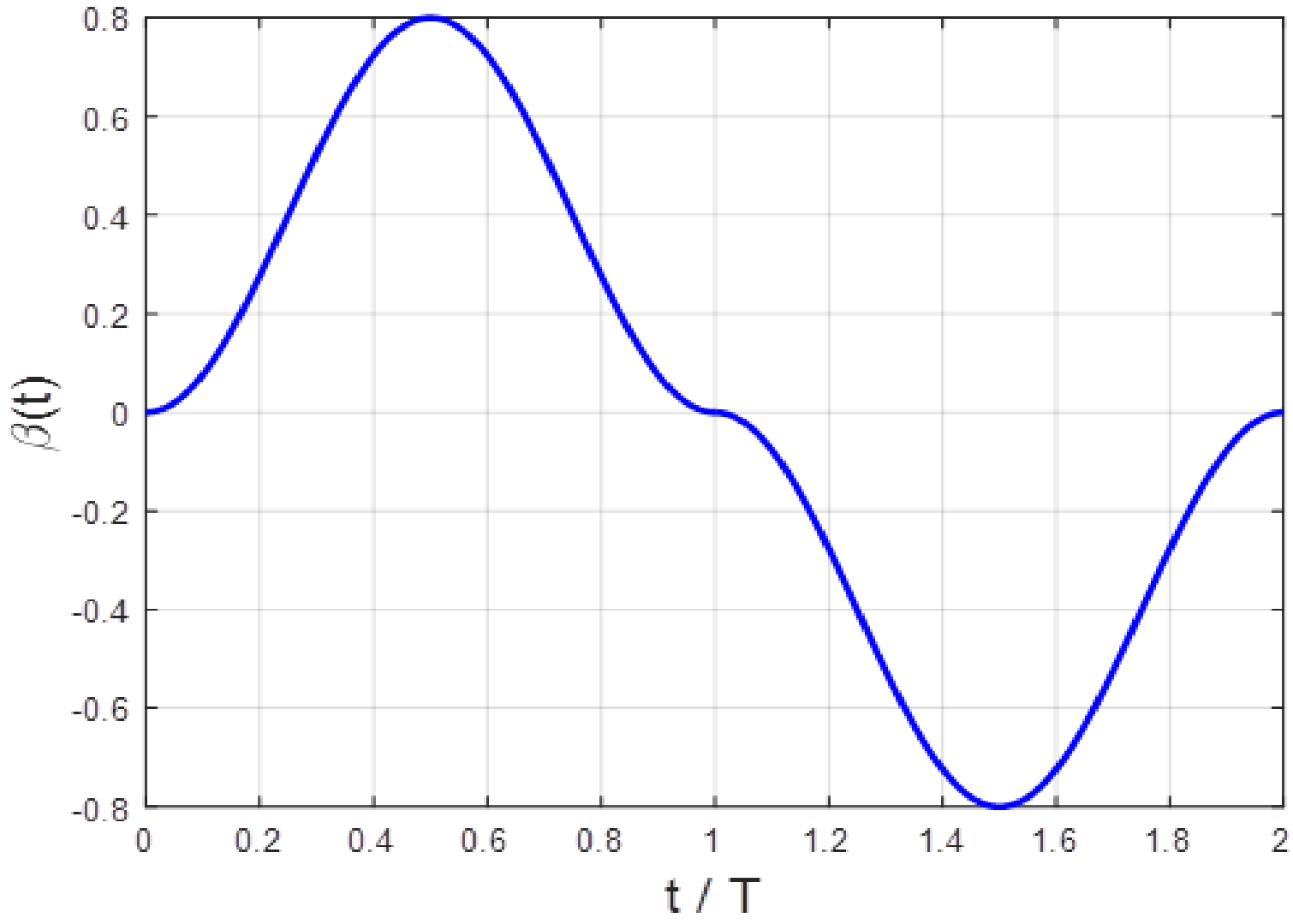

Fig. 4. Velocity profile for sinusoidal model ($\beta = 0.8$).

The proper-time interval of $\mathfrak{B}$ in terms of $dt$ is $dt' = dt\sqrt{1-\beta^2(t')}$. Upon formal integration via (1) one obtains an elliptic function:

$$t' = \int_0^t dt_1[1 - (\beta_0^2 \sin^2(\pi t/T)]^{1/2} = E\left(\tfrac{\pi t}{T} \backslash \arcsin \beta_0\right) \tag{7}$$

as given by Abramowitz & Stegun[11]. The results of numerical integration are plotted below, for $\beta_0 = 0.8,\ \ 0.6$, and 0.4 in Fig. 5, and they show a monotonic lag of $\mathfrak{B}$'s simultaneous time $t' = t_B$ behind the measured time $t = t_C$ of $\mathfrak{A}$.

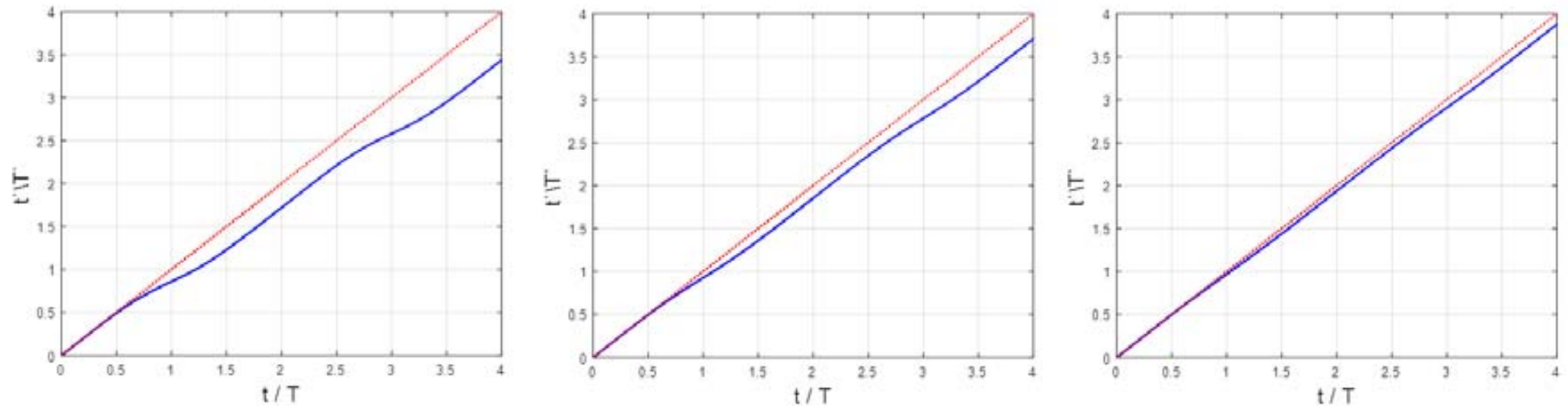

Fig. 5. Normalized time $t'(t)/T'$ (solid curve) vs. $t/T$ for $\beta_0 = 0.8$, 0.6, and 0.4.

Next, consider the situation from the point of view of traveler $\mathfrak{B}$ to find a counterpart to Fig. 3. First apply Eq. (4) for $t' < T'$. For $T' < t' < 2T'$, one needs to work out (4) using $\beta(t_1) < 0$ whenever $t_1 > T'$, i.e.

$$t = \int_{0'}^{T'} dt_1 [1 - \beta_1^2(t_1)])^{-1/2} + |\beta_1(t')| \int_{0'}^{T'} dt_1 \frac{\beta_1(t_1)}{\sqrt{1-\beta^2(t_1)}}$$
$$+ \int_{T'}^{t'} dt_1 [1 - \beta_1^2(t_1)])^{-1/2} - \beta_1(t') \int_{T'}^{t'} dt_1 \frac{\beta_1(t_1)}{\sqrt{1-\beta_1^2(t_1)}} \qquad (8)$$

Below are plots of Eq. $(8)$, for $\beta = 0.8$, $0.6$, and $0.4$ in Fig. $6$, in which $\mathfrak{A}$'s registered time $t$, simultaneous with the time $t'$ on $\mathfrak{B}$'s clock, is plotted as a function of $t'/T'$. The transition around the time $t' = T'$ now is gradual instead of abrupt as in Fig. 3. The same data are shown in Fig. 7 with $(t - t')/T$ plotted as a function of $t'/T'$.

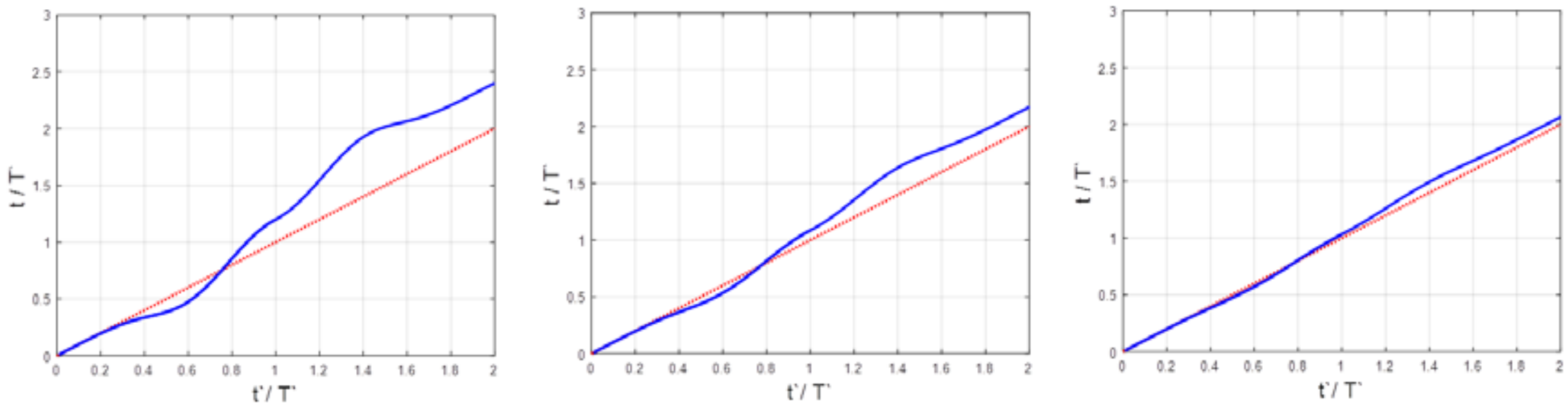

Fig. 6. Normalized $t/T$ (solid curve) as a function of $t'/T'$ for $\beta_0 = 0.8$, $0.6$, and $0.4$.

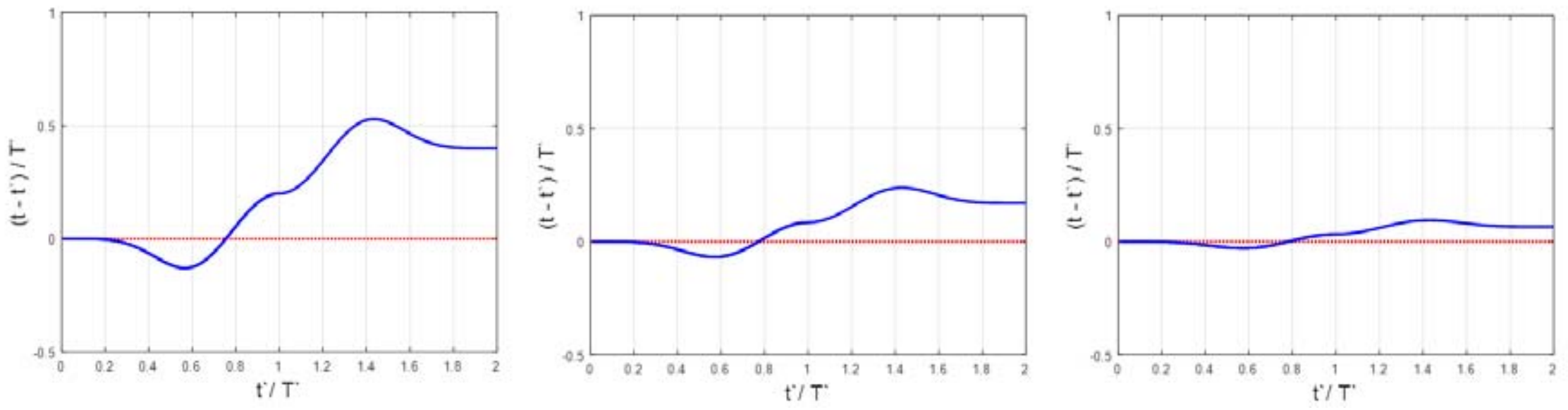

Fig. 7. $(t-t')/T$ (solid curve) as a function of $t'/T'$ for $\beta_0 = 0.8,\ 0.6$, and $0.4$.

The discontinuity of Fig. 3 is no longer present; instead, the 'aging' of $\mathfrak{A}$ is smooth and continuous, and it is centered around the midpoint of the two-way trip. It again more than compensates for the initial and final lagging of $\mathfrak{A}$'s clock behind $\mathfrak{B}$'s clock

## V. UNIFORM ACCELERATION TREATED IN STR

Another version of the TP will provide a useful test of Eq. (5). It is inspired by a problem in a number of articles[7,12-19]; one in which $\mathfrak{B}$ experiences uniform acceleration, e.g. $g \approx 9.8$ m/s$^2$ for $0 < t < T/2$, starting from $\beta(t=0) = 0$, followed by $-g$ for $T/2 < t < T$ at which time $\mathfrak{B}$ arrives at the destination star with zero velocity and acceleration. The return trip is symmetric for $T < t < 2T$. If indeed the coordinate frame of $t'$ moves with $\mathfrak{B}$, then $v' = 0$, and because the acceleration is then equivalent with gravitationally-induced spacetime curvature experienced by $\mathfrak{B}$ one might conclude that this problem requires general relativity (GTR). But this is not the case and we wish to show here that a result for $t(t')$ follows directly from the STR Eq. (5). It will then be demonstrated in section VI that the same result was obtained from the general theory by Perrin[7], and shown explicitly in a different form [namely as $t'(t)$] by Blecher[18, 19].

The first problem is that application of (4) requires knowledge of $\beta(t')$ which has to be obtained indirectly from the acceleration experienced by $\mathfrak{B}$ (this is worked out in Appendix 1)

$$\beta(t') = \tanh(gt'/c) \quad \text{for } t' < T'/2$$
$$\beta(t') = \tanh[g(T'-t')/c] \quad \text{for } T'/2 < t' < T' \tag{9}$$

The length of the trip is determined by the magnitude of $T'$; large values of $T'$ also indicate velocities close to $c$ for some interval around time $T'/2$.

The dependence $t_B(t_A)$ or $t'(t)$ is easily established (see Appendix 2) :

$$t' = \frac{c}{g}\text{arcsinh}\,\frac{gt}{c} \quad \text{for } 0 < t < T/2$$
$$\delta t' = \frac{c}{g}[\text{arcsinh}\,\frac{gT}{2c} + \text{arcsinh}\,\frac{g(t-T)}{c}] \quad \text{for } T/2 < t < T \tag{10}$$

Hence at the star, $t = T$ and $\delta t' = \frac{c}{g}\text{arcsinh}\,\frac{gT}{2c}$ which results in $t' = 2\frac{c}{g}\text{arcsinh}\,\frac{gT}{2c}$ upon return to Earth[18].

It requires appreciably more work to establish (see Appendix 3) $t(t')$: the time predicted by 𝔅 for 𝔄's clock from (4), namely $t_A(t_B)$, as

$$t = \frac{c}{g}\tanh(gt'/c) \qquad \text{for } t' < T'$$
$$t = \frac{c}{g}\left\{2\sinh\left(\frac{gT'}{2c}\right) - \tanh\left(\frac{g(T'-t')}{c}\right)\left[2\cosh\left(\frac{gT'}{2c}\right) - 1\right]\right\} \quad \text{for } T'/2 < t' < T' \tag{11a}$$

The two expressions are continuous at $t' = T'/2$. Upon arrival at the star ($t' = T'$), the second of these equations becomes $t = T = 2\frac{c}{g}\sinh\left(\frac{gT'}{2c}\right)$, same as in the line under (11). The prediction must agree with 𝔄's clock (as it indeed does) because 𝔄's clock can be synchronized with another on the star (which is in the same inertial system), and that other clock is then at the same location as 𝔅. The return trip is symmetric and therefore adds (12a) to $t = 2\frac{c}{g}\sinh\left(\frac{gT'}{2c}\right)$ for $T' < t' < 3T'/2$ (for the first line) and for $3T'/2 < t' < 2T'$ (for the second line). Therefore $t = 2T = 4\frac{c}{g}\sinh\left(\frac{gT'}{2c}\right)$ upon 𝔅's return to Earth[17]. Hence

$$t = 2\frac{c}{g}\sinh\left(\frac{gT'}{2c}\right) + \frac{c}{g}\tanh[\frac{g(t'-T')}{c}] \quad \text{for } T' < t' < 3T'/2$$
$$t = \frac{c}{g}\left\{4\sinh\left(\frac{gT'}{2c}\right) - \tanh\left(\frac{g[T'-(t'-T')]}{c}\right)\left[2\cosh\left(\frac{gT'}{2c}\right) - 1\right]\right\} \quad \text{for } 3T'/2 < t' < 2T' \tag{11b}$$

Figures 8 show plots of $t/T'$ vs. $t/T'$ for $T' = 2, 3,$ and $4$ y (top row) and $T' = 5,\ 8,$ and 10 y (bottom row). Results for $T < t < 2T$ need not be shown, because the return trip is symmetric and only adds on to the maximal value of $t(T')$.

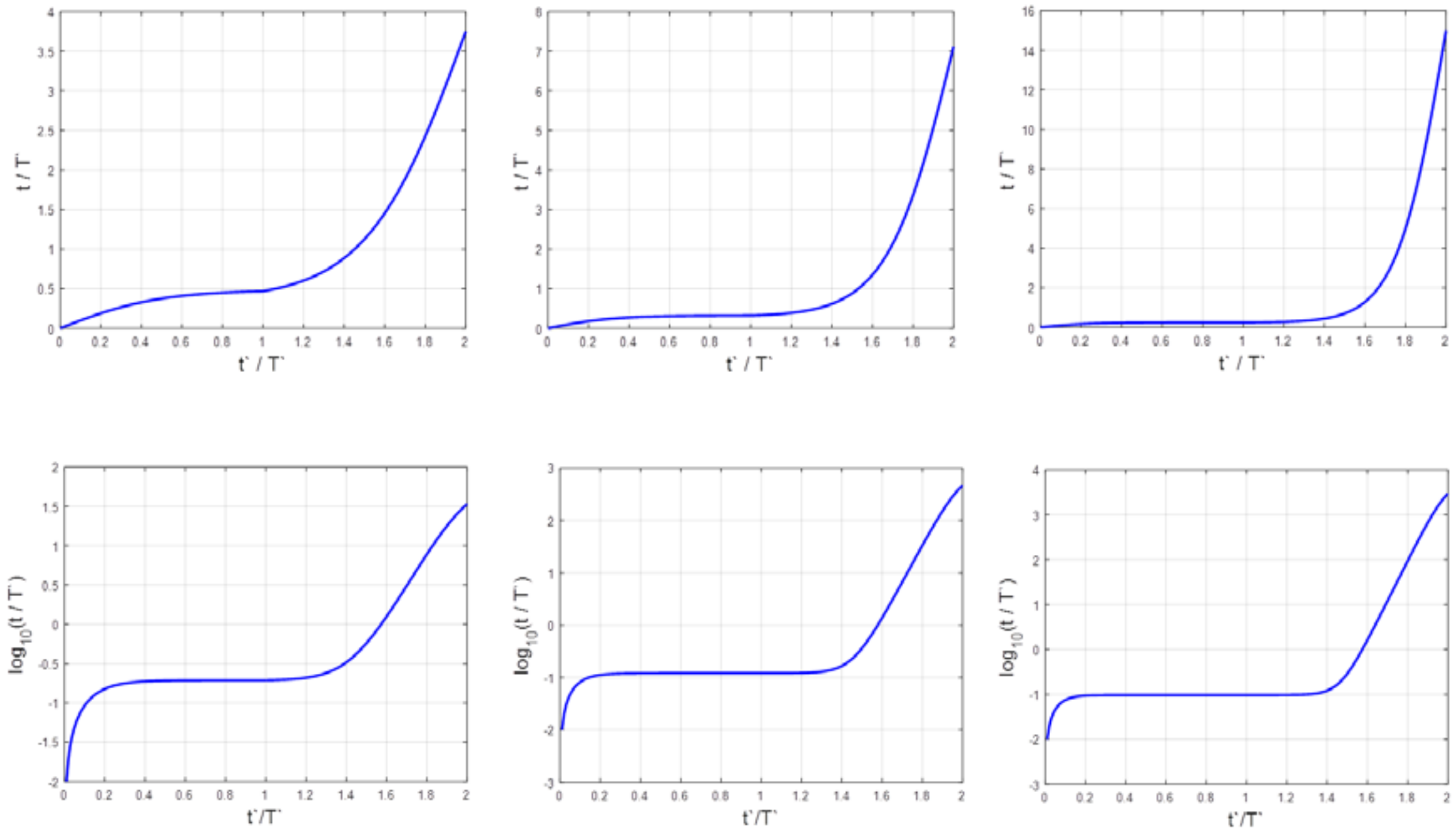

Fig. 8. $\mathfrak{B}$'s estimate of $\mathfrak{A}$'s clock time $t/T'$ vs. $t'/T'$ for $T' = 2,\ 3,$ and $4$ y (upper three) and
$T' = 5,\ 8,$ and $10$ y (lower three).

Whereas plots of $t(t')$ for the standard TP show a vertical jump in $t$ at $t' = T'$, the situation here is quite different; there appears to be an almost constant value of $t$ for a stretch of $t'$ around $t' = T'/2$ (when $T'$ is significantly larger than $1$. This appears to be due to the fact that $\beta(t')$, as given in (4), has an almost constant value near $\beta = 1$ in that stretch. The behavior of $t = t_A$, as a function of $t' = t_B$, can vary in different ways, depending upon the profile of $\beta(t')$.

Similar results were previously found by Müller et al[17]. Here, we've shown, in contrast, that the results can be obtained directly for a general equation, (5).

## VI. UNIFORM ACCELERATION TREATED IN GTR

In the previous section, acceleration is felt by $\mathfrak{B}$ as due to gravitationally-induced space curvature. Perrin[7] treats this with special relativity but also with general relativity (using a non-flat metric). Blecher[19] has attempted to translate Perrin's general-relativity result into an explicit dependence of $t = t_A$ upon $t' = t_B$ but his expression is in the form $t'(t)$ whereas a comparison with the result in section IV requires $t(t')$. His expression for $t'(t)$ is

$$t' = T' - \frac{c}{2g}\ln\left\{\frac{1-\beta_m}{1+\beta_m}\,\frac{B-(gt/c-\beta)}{-A+(gt/c-\beta)}\right\},\;\; \beta_m = \tanh(gT'/2c)$$
$$A = -2\sqrt{\frac{1-\beta_m}{1+\beta_m}} + (1-\beta_m), \quad B = 2\sqrt{\frac{1+\beta_m}{1-\beta_m}} - (1+\beta_m) \tag{12}$$

Inversion of this equation yields

$$t = \frac{c}{g}\left[1 + 2\frac{Q(\xi-1)-\xi}{Q+\xi^2}\right] \text{ given that } \xi \equiv \sqrt{\frac{1+\beta}{1-\beta}}, \quad Q \equiv e^{g(t'-T')/c} \tag{13}$$

Equation (13) can be shown (see Appendix 4) to be identical with (11a), and therefore confirms that the methods of sections II and IV lead to identical results for $t(t')$.

## VII. CONCLUSION

The fact that the proper time $t'(t)$ of an accelerating particle can be treated within STR was noted, among others, by Hartle[20]. That the effect is real has been verified by experiment[21,22]. What is new in this work is a generic equation for the time $t(t')$ that an accelerating observer predicts at his clock time $t'$ for the clock time $t$ in an inertial system, valid for any velocity distribution $v(t)$. A picture of the aging effect of $\mathfrak{A}$ also can be found for the standard twin paradox (no acceleration except at the initial, turn-around, and final points) in Muller[23]. But there are authors[24] who dispute the existence of a twin paradox; however that is not an opinion shared by most others.

### APPENDIX 1. DERIVATION OF EQS. (9)

Observer $\mathfrak{B}$ has velocity $v(t)$ as could be (but is not directly) observed in $\mathfrak{A}$'s inertial system $S$. $\mathfrak{B}$'s acceleration $g$ is experienced as gravitationally induced spacetime curvature. In $\mathfrak{A}$'s frame $S$:

4-vector $\boldsymbol{u} = \gamma(c, v, 0, 0)$ with $\gamma = (1 - v^2/c^2)^{-1/2} \equiv (1-\beta^2)^{-1/2}$
4-vector $\boldsymbol{a} = \frac{d\boldsymbol{u}}{d\tau} = \frac{dt}{d\tau}\frac{d\boldsymbol{u}}{dt} = \gamma\frac{d}{dt}[\gamma(c, v, 0, 0)] = (c\gamma\frac{d\gamma}{dt}, \gamma\frac{d(\gamma v)}{dt}, 0, 0)$

This yields: $a^0 = c\gamma\frac{d\gamma}{dt} = c\gamma^4\beta\frac{d\beta}{dt}$ and $a^1 = c\gamma^4\frac{d\beta}{dt}$ (A1)

Now perform the Lorentz transform of the 4-acceleration components using ${a'}^1 \equiv g$ in $\mathfrak{B}$'s frame for $0 < t < T/2$: With (A1) and $g \equiv {a'}^1 = \gamma(a^1 - \beta a^0)$ one obtains

$$g = c\gamma^3\frac{d\beta}{dt},\ \ \frac{d\beta}{dt} = \frac{g}{c}\gamma^{-3}\ \text{ leads to } dt = \frac{c}{g}(1-\beta^2)^{-3/2}d\beta = \frac{c}{g}d\Big[\beta(1-\beta^2)^{-1/2}\Big]$$
$$\text{and therefore }\ t - t_0 = \frac{c}{g}\Big[\beta(1-\beta^2)^{-1/2} - \beta_0(1-\beta_0^2)^{-1/2}\Big]$$

It is given that $\beta_0 = 0\,, t_0 = 0$: $\quad t = \frac{c}{g}\beta(1-\beta^2)^{-1/2},\quad \beta(t) = \frac{gt/c}{\sqrt{1+(gt/c)^2}}$ (A2)

Furthermore, Eq. (1) yields

$$t' = \int_0^t dt_1\Big\{1 - \frac{(gt_1/c)^2}{1+(gt_1/c)^2}\Big\}^{1/2} = \int_0^t dt_1[1+(gt_1/c)^2]^{-1/2} = \frac{c}{g}\text{arcsinh}\,\frac{gt}{c}\ \text{ resulting in}$$

$$t = \frac{c}{g}\sinh\frac{gt'}{c} \rightarrow \beta(t') = \frac{\sinh(gt'/c)}{\sqrt{1+\sinh^2(gt'/c)}} = \tanh(gt'/c) \qquad \text{(A3)}$$

(as also found by Müller, et al.[17]). Substitution into (A2) yields the first of Eqs. (10). Next, assume ${a'}^1 = -g$ for $T/2 < t < T$ for the second leg to the destination to obtain

$$\int_t^T dt_1 = \frac{c}{g}\int_\beta^{\beta_f} d\Big[\beta_1(1-\beta_1^2)^{-1/2}\Big]\ \text{ resulting in}$$
$$T - t = \frac{c}{g}\Big[\beta_f(1-\beta_f^2)^{-1/2} - \beta(1-\beta^2)^{-1/2}\Big]\ \text{and we may set } \beta_f = \beta(T) = 0$$
$$T - t = -\frac{c}{g}\beta(1-\beta^2)^{-1/2}\ \text{so that }\ \beta(t) = \frac{g(T-t)/c}{\sqrt{1+[g(T-t)/c]^2}} \qquad \text{(A4)}$$

which gives the correct (and continuous) result at $t = T/2$. In addition to these results, the time $t' = t_B$ of $\mathfrak{B}$ simultaneous with $\mathfrak{A}$'s clock time $t = t_C$ is then

$$T' - t' = \int_t^T dt_1\Big\{1 - \frac{[g(T-t_1)/c]^2}{1+[g(T-t_1)/c]^2}\Big\}^{-1/2} = \int_t^T dt_1\{1+[g(T-t_1)/c]^2\}^{-1/2}$$
$$= -\int_{T-t}^0 dt_1\big[1+[(gt_2/c)^2\big]^{-1/2} = \frac{c}{g}\text{arcsinh}\,\frac{g(T-t)}{c} \rightarrow$$
$$\frac{g(T-t)}{c} = \sinh\frac{g(T'-t')}{c} \qquad \text{(A5)}$$

This then yields the second of Eqs. (10) when substituted into (A4).

## APPENDIX 2. DERIVATION OF EQS. 10

The dependence of time $t' = t_B$ upon 𝔄's time $t = t_C$ follows directly from (A1) with application of (A2) and (A4):

$$t' = \int_0^t dt_1 [1-\beta^2(t_1)]^{1/2} \text{ and } dt_1 = \tfrac{c}{g}(1-\beta^2)^{-3/2} d\beta \rightarrow t' = \tfrac{c}{g}\int_0^\beta d\beta_1 (1-\beta_1^2)^{-1}$$

$$t' = \tfrac{c}{g}\text{arctanh}\,\beta(t) = \tfrac{c}{g}\text{arctanh}\,\tfrac{gt/c}{\sqrt{1+(gt/c)^2}} = \tfrac{c}{g}\text{arcsinh}\,\tfrac{gt}{c} \quad \text{for } 0 < t < T \qquad \text{(A6)}$$

For the 2nd leg, $T/2 < t < T$ one has acceleration $-g$ with added time $\delta t'$ beyond $T/2$:

$$\delta t' = \int_{T/2}^t dt_1 [1-\beta^2(t_1)]^{1/2} \text{ and } dt_1 = -\tfrac{c}{g}(1-\beta^2)^{-3/2} d\beta \text{ from which is obtained}$$

$$\delta t' = -\tfrac{c}{g}\int_{\beta(T/2)}^{\beta(t)} d\beta_1 (1-\beta_1^2)^{-1} = \tfrac{c}{g}[\text{arctanh}\,\beta(T/2) - \text{arctanh}\,\beta(t)]$$

$$= \tfrac{c}{g}\text{arcsinh}\,\tfrac{gT}{2c} + \tfrac{c}{g}\text{arcsinh}\,\tfrac{g(t-T)}{c} \text{ for } T/2 < t < T \qquad \text{(A7)}$$

## APPENDIX 3. DERIVATION OF EQS. 11A

For $t' < T'/2$ one finds from Eqs. (A2) and (A3)

$$[1-\beta^2(t_1)])^{1/2} = \left\{1 - \left[\tfrac{gt/c}{\sqrt{1+(gt/c)^2}}\right]^2\right\}^{-1/2} = \left[1 + (\tfrac{gt}{c})^2\right]^{1/2}$$

$$= \left[1 + \sinh^2 \tfrac{gt'}{c}\right]^{1/2} = \cosh\left(\tfrac{gt'}{c}\right)$$

$$\int_0^{t'} dt_1 [1-\beta^2(t_1)])^{1/2} = \int_0^{t'} dt_1 \cosh(\tfrac{gt'}{c}) = \tfrac{c}{g}\sinh(\tfrac{gt'}{c}) \qquad \text{(A8)}$$

$$\tfrac{\beta(t_1)}{\sqrt{1-\beta^2(t_1)}} = \tanh(\tfrac{gt_1}{c})\cosh(\tfrac{gt_1}{c}) = \sinh(\tfrac{gt_1'}{c})$$

$$\beta(t')\int_0^{t'} dt_1 \tfrac{\beta(t_1)}{\sqrt{1-\beta^2(t_1')}} = \tanh(\tfrac{gt'}{c})\int_0^{t'} dt_1 \sinh(g\tfrac{gt'}{c})$$

$$= \tfrac{c}{g}\tanh(\tfrac{gt'}{c})\left[\cosh(\tfrac{gt'}{c}) - 1\right] = \tfrac{c}{g}[\sinh(\tfrac{gt'}{c}) - \tanh(\tfrac{gt'}{c})] \qquad \text{(A9)}$$

This, when applied to (4) for $0 < t' < T'/2$, results in the first of Eqs. (11a). When the second of Eqs. (4) is applied for $T'/2 < t' < T'$ one finds similarly

$$[1-\beta^2(t_1)]^{-1/2} = \left[1 + \sinh\left(\tfrac{g(T'-t_1)}{c}\right)\right]^{-1/2} = \cosh\left(\tfrac{g(T'-t_1)}{c}\right)$$

$$\int_0^{t'} dt' [1-\beta^2(t_1)]^{-1/2} = \int_0^{T'/2} dt_1 \cosh(\tfrac{gt_1}{c}) + \int_{T'/2}^{t'} dt_1 \cosh\left(\tfrac{g(t_1-T')}{c}\right)$$

$$= \tfrac{c}{g}\left[\sinh\left(\tfrac{gT'}{2c}\right) + \sinh\left(\tfrac{g(t'-2')}{c}\right) - \sinh\left(\tfrac{-gT'}{2c}\right)\right] = \tfrac{c}{g}\left[2\sinh\left(\tfrac{gT'}{2c}\right) - \sinh\left(\tfrac{g(T'-t')}{c}\right)\right] \qquad \text{(A10)}$$

$$\int_0^{t'} dt_1 \beta(t_1)[1-\beta^2(t_1)]^{-1/2}$$

$$= \int_0^{T'/2} dt_1 \tanh\left(\tfrac{gt_1}{c}\right)\cosh\left(\tfrac{gt_1}{c}\right) + \int_{T'/2}^{t'} dt_1 \tanh\left(\tfrac{g(T'-t_1)}{c}\right)\cosh\left(\tfrac{g(T'-t_1)}{c}\right)$$
$$= \tfrac{c}{g}\left[\cosh\left(\tfrac{gT'}{2c}\right) - 1\right] + \tfrac{c}{g}\left\{\cosh\left(\tfrac{gT'}{2c}\right) - \cosh\left(\tfrac{g(T'-t')}{c}\right)\right\}$$
$$= \tfrac{c}{g}\left[2\cosh\left(\tfrac{gT'}{2c}\right) - \cosh\left(\tfrac{g(T'-t')}{c}\right) - 1\right] \quad \text{(A11)}$$

$$\beta(t' \int_0^{t'} dt_1' \beta(t_1)[1-\beta^2(t_1)]^{-1/2}$$
$$= \tfrac{c}{g}\tanh\left(\tfrac{g(T'-t')}{c}\right)\left[2\cosh\left(\tfrac{gT'}{2c}\right) - \cosh\left(\tfrac{g(T'-t')}{c}\right) - 1\right]$$
$$= \tfrac{c}{g}\left[2\cosh\left(\tfrac{gT'}{2c}\right) - 1\right]\tanh\left(\tfrac{g(T'-t')}{c}\right) - \tfrac{c}{g}\sinh\left(\tfrac{g(T'-t')}{c}\right) \quad \text{(A12)}$$

Together this yields the desired result.

$$t = \tfrac{c}{g}\left[2\sinh\left(\tfrac{gT'}{2c}\right) - \sinh\left(\tfrac{g(T'-t')}{c}\right)\right]$$
$$- \tfrac{c}{g}\left[2\cosh\left(\tfrac{gT'}{2c}\right) - 1\right]\tanh\left(\tfrac{g(T'-t')}{c}\right) + \tfrac{c}{g}\sinh\left(\tfrac{g(T'-t')}{c}\right)$$
$$= \tfrac{c}{g}\left\{2\sinh\left(\tfrac{gT'}{2c}\right) - \tanh\left(\tfrac{g(T'-t')}{c}\right)\left[2\cosh\left(\tfrac{gT'}{2c}\right) - 1\right]\right\} \quad \text{(A13)}$$

The derivation for $T' < t' < 2T'$ is essentially the same, except for minor details.

## APPENDIX 4. EQUIVALENCE OF EQS. (12) AND (14)

Here, it is demonstrated briefly that Eqs. (13) and (11a) are identical. Starting from (11a), the following algebraic changes lead to the desired result:

$$\tfrac{gt}{c} = \left\{2\sinh\left(\tfrac{gT'}{2c}\right) - \tanh\left(\tfrac{g(T'-t')}{c}\right)\left[2\cosh\left(\tfrac{gT'}{2c}\right) - 1\right]\right\}$$
$$= \left(e^{gT'/2c} - e^{-gT'/2c}\right) - \tfrac{e^{g(T'-t')/c} - e^{-g(T'-t')/c}}{e^{g(T'-t')/c} + e^{-g(T'-t')/c}}\left[e^{gT'/2c} + e^{-gT'/2c} - 1\right]$$
$$= \left(e^{gT'/2c} - e^{-gT'/2c}\right) - \tfrac{e^{gT'/c} - e^{-g(T'-t')/c}}{e^{gT'/c} + e^{-g(T'-t')/c}}\left[e^{gT'/2c} + e^{-gT'/2c} - 1\right] \quad \text{(A14)}$$

Define $\xi \equiv \sqrt{\frac{1+\beta}{1-\beta}} = e^{gT'/2c}$ [which follows from (A2) and (A3)] and $Q = e^{g(t'-T')/c}$, and the above becomes

$$\tfrac{gt}{c} = \left(\xi - \tfrac{1}{\xi}\right) - \tfrac{\xi^2 - Q}{\xi^2 + Q}\left(\xi + \tfrac{1}{\xi} - 1\right) = \frac{(\xi^2+Q)\left(\xi - \frac{1}{\xi}\right) - (\xi^2 - Q)\left(\xi + \frac{1}{\xi} - 1\right)}{\xi^2 + Q} \quad \text{(A15)}$$

This then yields $\frac{gt}{c} = \frac{2\xi(Q-1) + (\xi^2 - Q)}{\xi^2 + Q} = 1 + \frac{2\xi(Q-1) + (\xi^2 - Q) - (\xi^2 + Q)}{\xi^2 + Q} = 1 + 2\frac{Q(\xi - 1) - \xi}{\xi^2 + Q}$
which is Eq. (13).

## ACKNOWLEDGMENT

I am indebted to M. Blecher for many discussions, and for making Eq. (12) available to me.

*Emeritus professor Bradley Department of Electrical & Computer Engineering*

## REFERENCES

1 J. B. Hartle, *Gravity, an introduction to Einstein's general relativity,* Pearson Education, UK, 2003, sections 4.4 [for Eq. (1)] and 4.3 (for hyperbolic geometry).

2 T. Takeuchi, *Special Relativity*, Encyclopedia of Applied High Energy and Particle Physics, pp. 47-94, 2009, Wiley VCH.

3 R. H. Romer, "Twin paradox in special relativity," Am. J. Phys. **27**, 1959, pp. 131-135.

4 R. A.Muller, "The twin paradox in special relativity," Am. J. Phys. **40**, 1972, pp. 966-969.

5 M. Sachs, "A resolution of the clock paradox," Physics Today **24**, 1972, pp. 23-29.

6 T. Dray, "The twin paradox revisited," Am. J. Phys. **58**, 1990, pp. 822-825.

7 R. Perrin. "Twin paradox: A complete treatment from the point of view of each twin," Am. J. Phys. **47**, 1979, pp. 317-319.

8 D. A. de Wolf, "Aging and communication in the Twin Paradox," Eur. J. Phys., **37**, 2016.

9 A. Einstein, *Zur Elektrodynamik bewegter Körper,* Annalen der Physik 17, 1905, pp. 891-921.

10 A similar formula, Eq. (3) in ref. 8, contains a typographical error and is not complete.

11 M. Abramowitz and I. A. Stegun, *Handbook of Mathematical Functions,* NBS Applied Mathematics Series 55, June 1964, Equation 17.2.9.

12 E. Eriksen and O. Grøn, "Relativistic dynamics in uniformly accelerated reference frames with application to the clock paradox," Eur. J. Phys. **11**, 1990, pp. 39-44.

13 L. Iorio, "On the clock paradox in the case of circular motion of the moving clock." Eur. J. Phys. **26**, 2005, pp. 535-541.

14 S. Boblest, T Müller, and G. Wunner, "Twin paradox in de Sitter spacetime," Eur. J. Phys. **32**, 2011.

15 M. Blecher, General Relativity, a first examination, World-Scientific, Singapore, 2016

[16] M. Gasperini, "The twin paradox in the presence of gravity," Mod. Phys. Lett. A29, 2014, 1450149.

[17] T. Müller, A. King, and D. Adis, " A trip to the end of the universe and the twin "paradox"," Am. J. Phys. **76**, 2008, pp. 360-373.

[18] M. Blecher, General Relativity, a first examination, World-Scientific, Singapore, 2016

[19] M. Blecher, private communication.

[20] J. B. Hartle, *Op. cit.,* Eq. (4.14)

[21] J. C. Hafele and R. E. Keating, "Around-the-World Atomic Clocks: Predicted Relativistic Time Gain," Science **177** (4044), 166-168 (1972).

[22] L. J. Wang, "Symmetric experiments to test the clock paradox," Physics and Modern Topics in Mechanical and Electrical Engineering, (ISBN 960-8052-10-6, 1999), pp. 45-50.

[23] R. A.Muller, "The twin paradox in special relativity," Am. J. Phys. **40**, 1972, pp. 966-969.

[24] M. Harada, "The twin paradox repudiated," Physics Essays **24**, 2011, pp. 454-455.